\documentclass[twocolumn]{revtex4}
\usepackage{amsmath,amssymb,dsfont,graphicx,times}

\renewcommand{\S}{section }

\begin{document}
\title{From scattering theory to complex wave dynamics in
non-hermitian $\mathcal{PT}$-symmetric resonators}
\author{Henning Schomerus}
\affiliation{Department of Physics, Lancaster University, Lancaster, LA1 4YB, United Kingdom}
\date{\today}
\begin{abstract}I review how methods from mesoscopic physics can be applied to
describe the multiple wave scattering and complex wave dynamics in
non-hermitian $\mathcal{PT}$-symmetric resonators, where an absorbing
region is coupled symmetrically to an amplifying region. Scattering
theory serves as a convenient tool to classify the symmetries beyond the
single-channel case and leads to effective descriptions which can be
formulated in the energy domain (via Hamiltonians) and in the time domain
(via time evolution operators). These models can then be used to identify
the mesoscopic time and energy scales which govern the spectral
transition from real  to complex eigenvalues. The possible presence of
magneto-optical effects (a finite vector potential) in multichannel
systems leads to a variant (termed $\mathcal{PTT}'$ symmetry) which
imposes the same spectral constraints as $\mathcal{PT}$ symmetry. I also
provide multichannel versions of generalized  flux-conservation laws. \end{abstract}

\maketitle

\section{\label{sec:1}Introduction and Motivation}
Optical systems combining lossy and active elements provide a platform to
implement analogues of non-hermitian $\mathcal{PT}$-symmetric quantum
systems \cite{bender,ptreview1,ptreview2}, which allows to realize unique
optical switching effects
\cite{christodoulidesgroup1,christodoulidesgroup2,christodoulidesgroup3,
ptexperiments1,ptexperiments2,pteffects1,pteffects2a,pteffects4,pteffects5,
hsqoptics,longhilaserabsorber,variousstonepapers1,variousstonepapers2,yoo,schomerusyh}.
The wave scattering and dynamics in these systems provide an aspect which
also permeates mesoscopic phenomena in microcavity lasers \cite{vahala},
coherent electronic transport \cite{beenakkerreview,datta},
superconductivity \cite{beenakkerreview,altlandzirnbauer}, and
quantum-chaotic dynamics \cite{haake}. The main goal here is to relate
how tools established in these disciplines can be used to approach the
exciting spectral and dynamical features in $\mathcal{PT}$-symmetric
optics. Scattering theory \cite{beenakkerreview,datta,fyodorovsommers},
in particular, is ideally suited to deal with the complications of
complex wave dynamics in multichannel situations, which appear when
one goes beyond one-dimensional situations, and also easily accounts for
any leakage if a system is geometrically open, as is often required by
the nature of the desired optical effects, or because of the design of
practical devices.

We thus describe in detail how this approach can be applied to optical
realizations of $\mathcal{PT}$-symmetric systems with a wide range of
geometries \cite{hsqoptics,longhilaserabsorber,variousstonepapers1,
variousstonepapers2,yoo,hsrmt,birchallweyl,birchallrmt}. (For
$\mathcal{PT}$-symmetric scattering in effectively one dimension see,
\emph{e.g.}, Refs.~\cite{cannata,berry,jones}.) To formulate the approach
under these conditions it matters whether a geometric symmetry inverts or
preserves the handedness of the coordinate system, especially in the
presence of magneto-optical effects (vector potentials, which generally
cannot be gauged away beyond one dimension). An analogous bifurcation
arises in the specification of the time-reversal operation. Moreover, it
is important to note that the fields defining a device geometry are
external, a distinction from particle-physics settings which is reflected
in the subtleties of Onsager's reciprocal relations \cite{onsager}. Thus
the formulation of $\mathcal{PT}$ symmetry itself requires some care. In
particular, an alternative appears, termed $\mathcal{PTT}'$ symmetry
\cite{hsrmt}, which results in the same spectral constraints but imposes
different physical symmetry conditions. The scattering approach also
allows to restate the symmetry requirements in terms of generalized
conservation laws \cite{variousstonepapers2}, which we here extend to
multiple channels. Furthermore, the approach leads to effective models of
complex wave dynamics which relate the spectral features to universal
mesoscopic time and energy scales \cite{hsrmt,birchallweyl,birchallrmt}.

We start this exposition with a brief recapitulation of the analogy
between optics and non-hermitian quantum mechanics (\S\ref{sec:2}), and
discuss variants of geometric and time-reversal operations which  can be
used to set up $\mathcal{PT}$ and $\mathcal{PTT}'$ symmetry
(\S\ref{sec:3} and \S\ref{sec:4}). We then describe how scattering theory
can be applied to study the spectral features in these settings, first
generally (\S\ref{sec:5}-\S\ref{sec:7}) and then for a coupled-resonator
geometry (\S\ref{sec:8}). This leads to effective models (\S\ref{sec:9})
which can be formulated in the energy domain (via Hamiltonians) and in
the time domain (via time evolution operators), and capture the relevant
energy and time scales (\S\ref{sec:10}). The concluding \S\ref{sec:11}
also describes possible generalizations of these models.

\section{\label{sec:2}Optical analogues of non-hermitian quantum mechanics}

\begin{figure}[b]
\includegraphics[width=\columnwidth]{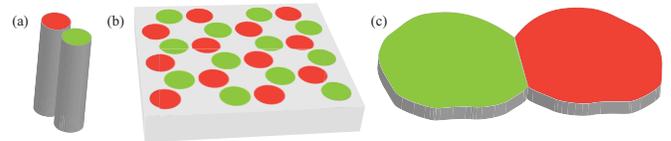}
\caption{\label{fig1}(Online version in colour.)
Common designs of $\mathcal{PT}$-symmetric optical systems with absorbing
(light green) and amplifying (dark red) elements. (a) Two optical fibers.
(b) Localized modes in a photonic crystals. (c) Coupled-resonator
geometry. }
\end{figure}

The $\mathcal{PT}$-symmetric optical systems mentioned in the
introduction cover a large range of designs, including coupled optical
fibres, photonic crystals and coupled resonators, as depicted in figure
\ref{fig1}. Some of the elements are absorbing while others are
amplifying, and the absorption and amplification  rates, geometry, and
other material properties are carefully matched to result in a symmetric
arrangement.

We concentrate on two common designs, effectively two-dimensional systems
(planar resonators or photonic crystals, with coordinates $x$, $y$) as
well as three-dimensional pillar-like systems (such as arrangements of
aligned  optical fibers with a fixed cross section) where the geometry
does not depend on the third coordinate $z$, and assume that the relevant
effects of wave scattering, gain and loss can be subsumed in a refractive
index $n(\mathbf{r})$, where ${\rm Im}\, n(\mathbf{r})>0$ signifies
absorption (loss) while ${\rm Im}\, n(\mathbf{r})<0$ signifies
amplification (gain). Later on, we also include an external vector
potential $\mathbf{A}(\mathbf{r})$ representing magneto-optical effects.

These assumptions allow to separate transverse magnetic (TM) and electric
(TE) modes, with the magnetic or electric field, respectively, confined
to the $xy$-plane (thus transverse to $z$). The electromagnetic field can
then be described by a scalar wave function $\psi({\bf r})$, which
represents the $z$-component of the electric or magnetic field,
respectively. One now arrives at two principal situations, which both
serve as an optical analogue of non-hermitian quantum mechanics.

\subsection{Helmholtz equation}

On the one hand, one can focus on the propagation in the $xy$-plane,
resulting in an effectively two-dimensional system which is described by
a Helmholtz equation
\begin{align}\label{eq:helmholtz}
&\mathcal{L}(\omega)\psi({\bf r})=0,
\nonumber \\
&\mathcal{L}(\omega)=
\left\{\begin{array}{cc}
\nabla^2+\frac{\omega^2n^2({\bf r})}{c^2} & \mbox{(TM)} \\
\nabla \frac{1}{n^{2}({\bf r})} \nabla +\frac{\omega^2}{c^2}& \mbox{(TE)}  \end{array}\right.
,\quad\nabla=\left(\begin{array}{c} \partial_x\\ \partial_y\end{array}\right)
.
\end{align}
The Helmholtz equation is analogous to a stationary Schr{\"o}dinger
equation, thus, an eigenvalue problem for the frequencies $\omega$, which
can become complex due to the loss and gain encoded in $n$, or because of
leakage at the boundaries of the system in the propagation plane. This
then  describes quasistationary states which decay (${\rm Im}\,\omega<0$)
or grow (${\rm Im}\,\omega>0$)  over time.

\subsection{Paraxial equation}

On the other hand, one can focus on the propagation into the
$z$-direction perpendicular to this plane, which is then often described
by a paraxial equation
\begin{align}
&2i\kappa\partial_z \psi({\bf r}) +\mathcal{L}(\kappa)\psi({\bf r})=0,
\nonumber \\ &
{\mathcal{L}}(\kappa)=
\left\{\begin{array}{cc}
\nabla^2+\frac{\omega^2n^2({\bf r})}{c^2} -\kappa^2& \mbox{(TM)} \\
n^2({\bf r})\nabla \frac{1}{n^{2}({\bf r})}  \nabla +\frac{\omega^2n^2({\bf r})}{c^2} -\kappa^2& \mbox{(TE)}  \end{array}\right. .
\label{eq:paraxial}
\end{align}
Here $\psi({\bf r})$ now is an envelope wave function, obtained after
separating out a term $\exp(i\kappa z)$. This equation remains valid if
$n$ varies only slowly with $z$, $|\partial_z n/n|\ll|\kappa|$.

The paraxial equation is an analogue of a time-dependent Schr{\"o}dinger
equation, thus, a dynamical equation where an initial condition is
propagated forwards---here, not in time, but along the spatial direction
with coordinate $z$. It is then natural to probe the system at an initial
and a final cross-section, while the physical frequency $\omega$ is now
real.

As we assume that $n$ is $z$-independent, the paraxial equation is
associated with the eigenvalue problem $\mathcal{L}(\kappa)\psi({\bf
r})=0$ for the propagation constant $\kappa$. This eigenvalue problem
shares all relevant mathematical features with the problem
\eqref{eq:helmholtz} for $\omega$ (in particular, the eigenvalues again
can be complex because of loss, gain, and leakage in the system); so do
variants that rely on wave localization in the cores of optical fibers or
Wannier-like modes in a photonic crystal, with the gradient replaced by
hopping terms. For definiteness, we shall employ notations adapted to the
eigenvalue problem for $\omega$.

\section{\label{sec:3}Symmetries of the wave equation}
The investigation of symmetries in higher-dimensional systems of
arbitrary geometry is a central aspect of mesoscopic physics. For
hermitian systems, a complete classification requires to take care of
magnetic fields, internal degrees of freedom such as spin, pairing
potentials and particle-hole symmetries
\cite{beenakkerreview,datta,altlandzirnbauer,haake}. Non-hermitian
systems provide an even wider setting, with a mathematical
classification, \emph{e.g.}, provided in Ref. \cite{magnea}. In order to
identify and specify the role of symmetries for their optical
realizations we denote the operator in the wave equation by
$\mathcal{L}(\omega; n(\mathbf{r}))$, which explicitly takes care of the
functional dependence on the refractive index $n$.

\subsection{$\mathcal{PT}$ symmetry}
In $\mathcal{PT}$-symmetric quantum mechanics
\cite{bender,ptreview1,ptreview2}, the parity operator $\mathcal{P}$
generally stands for a unitary transformation which squares to the
identity. In optical systems, this operation is usually realized
geometrically via an isometric involution, still denoted as
$\mathcal{P}$, which inverts one, two, or three coordinates. To specify
the consequences for the wave equation we promote ${\cal P}$ to a
superoperator (we employ this notation as we will also encounter a
transformation $\mathcal{T}'$ which does not correspond to an ordinary
unitary or antiunitary operator). Then
\begin{equation}\label{eq:p}
\mathcal{P}[\mathcal{L}(\omega; n(\mathbf{r}))]
=\mathcal{L}(\omega; n(\mathcal{P}\mathbf{r})),
\end{equation}
which should be read as a rule how to write the wave equation in the
transformed coordinate system. The transformed equation is then solved by
${\cal P}\psi({\bf r})\equiv\psi({\cal P}{\bf r})$.

Conventional time reversal is implemented by complex conjugation in the
position representation, which constitutes an antiunitary operation. We
then have
\begin{equation}
\mathcal{T}[\mathcal{L}(\omega; n(\mathbf{r}))]
=\mathcal{L}(\omega^*; n^*(\mathbf{r})),
\end{equation}
which delivers a wave equation solved by ${\cal T}\psi({\bf r})\equiv\psi^*({\bf r})$.

The wave equation now displays $\mathcal{PT}$ symmetry if the refractive
index obeys
$n({\bf r})=n^*({\cal P}{\bf r})$.
In this situation ${\cal P}$ is an involution that interchanges
amplifying and absorbing regions with matching amplification and
absorption rates, and
\begin{equation}
\mathcal{PT}[\mathcal{L}(\omega; n(\mathbf{r}))]
=\mathcal{L}(\omega^*; n^*(\mathcal{P}\mathbf{r}))
=\mathcal{L}(\omega^*; n(\mathbf{r})),
\end{equation}
which constraints the spectral properties of $\mathcal{L}$ if the
boundary conditions also respect the symmetry (this is typically
\emph{not} the case if the system is open). The eigenvalues $\omega_n$
then obey
\begin{equation}\label{eq:constraint}
\omega_n=\omega^*_{\bar n}
\end{equation}
and thus either real ($n=\bar n$) if $\psi^*_n(\mathcal{P}\mathbf{r})$
linearly depends on $\psi_n(\mathbf{r})$, or occur in complex-conjugate
pairs ($n\neq\bar n$) if that is not the case.

\subsection{$\mathcal{T'}$ symmetry}
It is important to distinguish the conventional antiunitary time-reversal
operation ${\cal T}$ from another operation that is also often termed
time-reversal \cite{beenakkerreview,datta,altlandzirnbauer,haake}. If
$\mathcal{L}({\bf r})$ were hermitian ($\omega$ and $n$ both real with
appropriate boundary conditions), then the action of ${\cal T}$ could not
be distinguished from the action of a superoperator ${\cal T}'$ acting in
the position representation as
\begin{equation}
{\cal T}'[\mathcal{L}(\omega; n(\mathbf{r}))]=\mathcal{L}^T(\omega; n(\mathbf{r})).
\end{equation}
Thus, ${\cal T}'$ transforms the right eigenvalue problem ${\cal
L}\psi=0$ into the left eigenvalue problem ${\cal L}^T \bar\psi=0$. This
delivers the same spectrum, and while the right and left eigenfunctions
$\psi$ and $\bar\psi$ for a given eigenvalue generally differ they are
related by biorthogonality constraints. In the presently assumed absence
of magneto-optical effects, ${\cal T}'$ is indeed an exact symmetry,
$\mathcal{L}^T(\omega; n(\mathbf{r})) =\mathcal{L}(\omega;
n(\mathbf{r}))$ and thus $\psi=\bar\psi$. This holds even if $n$ is
complex, thus, even if  ${\cal T}$ symmetry is broken. In the
non-hermitian case, therefore, these two operations are distinct and must
be treated separately.

\section{\label{sec:4}Magneto-optical effects}
In mesoscopic systems with complex wave dynamics, the alternative
time-reversal operation ${\cal T}'$ governs a multitude of effects
ranging from coherent backscattering and wave localization to minigaps in
mesoscopic superconductors
\cite{beenakkerreview,datta,altlandzirnbauer,melsen}. To clarify the role
of this operation we now consider magneto-optical effects, which are
described by a (possibly complex) external vector potential
$\mathbf{A}(\mathbf{r})$ that enters the wave equation through terms of
the generic form $(\nabla+i\mathbf{A}(\mathbf{r}))^2$. For this purpose
we denote the operator in the wave equation as $\mathcal{L}(\omega;
n(\mathbf{r}), \mathbf{A}(\mathbf{r}))$. As before, we continue to focus
on effectively two-dimensional systems; thus, $A_z=0$, see
\eqref{eq:helmholtz} and \eqref{eq:paraxial}.

\subsection{$\mathcal{PT}$ symmetry}
The involution $\mathcal{P}$ (interpreted as a  coordinate transformation
of the wave equation) transforms the external vector potential according
to
$(\nabla+i\mathbf{A}(\mathbf{r}))^2\to
({\cal P}\nabla+i\mathbf{A}({\cal P}\mathbf{r}))^2
=(\nabla+i[{\cal P}\mathbf{A}]({\cal P}\mathbf{r}))^2
$,
as ${\cal P}$ is an isometry.
Thus\begin{equation}
\mathcal{P}[\mathcal{L}(\omega; n(\mathbf{r}), \mathbf{A}(\mathbf{r}))]
=\mathcal{L}(\omega; n(\mathcal{P}\mathbf{r}),
[\mathcal{P}\mathbf{A}](\mathcal{P}\mathbf{r}) ).
\end{equation}
This analysis reveals an important feature of parity when applied to
external fields, which sets them apart from internal fields that are
often discussed in the setting of particle physics, and then are subject
to additional explicit transformation rules. Here we are interested in
symmetries of a wave function $\psi({\bf r})$ that only accounts for the
internal system dynamics, not for external components such as the motion
of electrons in the inductors or the magnetic moments in the permanent
magnets generating the field. This distinction is fundamental for
Onsager's reciprocal relations, which break down in the presence of
external magnetic fields \cite{onsager}. The resulting magnetic field
depends on whether $\mathcal{P}$ preserves or inverts the handedness of
the coordinate system, as
$\nabla\times\mathcal{P}\mathbf{A}(\mathcal{P}\mathbf{r})={\rm
det}(\mathcal{P})\mathcal{P}[\nabla \times \mathbf{A}(\mathbf{r})]$.

\begin{figure}
\includegraphics[width=\columnwidth]{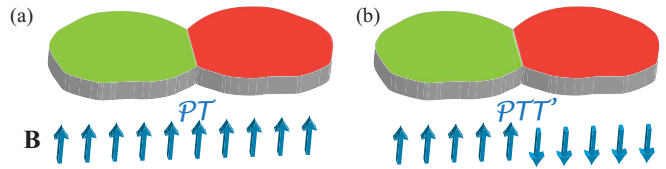}
\caption{\label{fig2}(Online version in colour.)
Illustration of magneto-optical effects (associated with an effective magnetic field
$\mathbf{B}$) for (a) $\mathcal{PT}$ symmetry and (b) $\mathcal{PTT}'$
symmetry, where $\mathcal{P}$ is a reflection \cite{hsrmt}. }
\end{figure}

The time-reversal operator $\mathcal{T}$ transforms
$(\nabla+i\mathbf{A}(\mathbf{r}))^2\to (\nabla-i\mathbf{A}^*(\mathbf{r}))^2
$,
so
\begin{equation}
\mathcal{T}[\mathcal{L}(\omega; n(\mathbf{r}), \mathbf{A}(\mathbf{r}))]
=\mathcal{L}(\omega^*; n^*(\mathbf{r}), -\mathbf{A}^*(\mathbf{r})).
\end{equation}
$\mathcal{PT}$ symmetry thus requires
\begin{equation}\label{eq:ptna}
n({\bf r})=n^*({\cal P}{\bf r}),\quad
\mathbf{A}({\bf r})=-\mathcal{P}\mathbf{A}^*({\cal P}{\bf r}).
\end{equation}
\emph{E.g.}, when ${\cal P}$ is a reflection $x\to-x$ this holds for a
homogeneous magnetic field that points in the $z$ direction [see figure
\ref{fig2}(a)].

\subsection{$\mathcal{PTT}'$ symmetry}
We now identify a modified symmetry, which eventually yields the same
spectral constraints as obtained for $\mathcal{PT}$ symmetry, but imposes
a different condition on the magnetic field \cite{hsrmt}.  This variant
follows from the inspection of the $\mathcal{T'}$ operation. In the
position representation $\partial_x^T=-\partial_x$,
$\partial_y^T=-\partial_y$ (recall that we focus on the effects in the 2D
cross-sectional plane), and thus
\begin{equation}
\mathcal{T'}:\quad (\nabla+i\mathbf{A}(\mathbf{r}))^2\to
[(\nabla+i\mathbf{A}(\mathbf{r}))^2]^T
=(\nabla-i\mathbf{A}(\mathbf{r}))^2
.
\end{equation}
This effectively inverts the vector potential,  \begin{equation}
{\cal T}'[\mathcal{L}(\omega; n(\mathbf{r}), \mathbf{A}(\mathbf{r}))]=
\mathcal{L}(\omega; n(\mathbf{r}), -\mathbf{A}(\mathbf{r})),
\end{equation}
which confirms that  ${\cal T}'$ symmetry is broken when $\mathbf{A}$ is
finite. Consider now
\begin{equation}
\mathcal{PTT'}[\mathcal{L}(\omega; n({\bf r}),\mathbf{A}({\bf r})]
=\mathcal{L}(\omega^*; n^*({\cal P}{\bf r}), \mathcal{P}\mathbf{A}^*({\cal P}{\bf r})).
\end{equation}
This turns into a symmetry if
\begin{equation}\label{eq:pttna}
n({\bf r})=n^*({\cal P}{\bf r}),\quad
\mathbf{A}({\bf r})=\mathcal{P}\mathbf{A}^*({\cal P}{\bf r}).
\end{equation}
\emph{E.g.}, when ${\cal P}$ is a reflection $x\to-x$ this holds for an
antisymmetric magnetic field which (in the plane of the resonator) points in the $z$ direction  [see figure
\ref{fig2}(b)].

\section{\label{sec:5}Scattering formalism}

The optical systems considered here are naturally open, with leakage
occurring, \emph{e.g.}, at the fiber tips and waveguide entries, or
because the confinement in the cross-section relies on partial internal
reflection at refractive index steps or semitransparent mirrors. These
systems can thus be probed via scattering
\cite{beenakkerreview,datta,fyodorovsommers}, which delivers
comprehensive insight into their spectral and dynamical properties and
illuminates the consequences of the symmetries discussed above, as well
as the role of multiple scattering addressed later on.

These properties are encoded in the scattering matrix $S(\omega)$, which
relates the amplitudes $a_{\mathrm{in},n}$, $a_{\mathrm{out},n}$ in
incoming and outgoing  scattering states $\chi_{\mathrm{in},n}$,
$\chi_{\mathrm{out},n}$,
\begin{equation}\label{eq:sdef}
\mathbf{a}_{\mathrm{out}}=S(\omega) \mathbf{a}_{\mathrm{in}}.
\end{equation}
This relation is generally obtained by the solution of the wave equation
under appropriate conditions at the boundary $\partial\Omega$ of the
scattering region $\Omega$ (outside of which we set $n=1$, ${\bf A}=0$).
The scattering states are  assumed to be flux orthonormalised,
\begin{equation}\label{eq:flux}
\int_{\partial \Omega} d\mathbf{S}\cdot
[\chi_{\sigma,n}^*\nabla \chi_{\sigma',m}- \chi_{\sigma',m}\nabla\chi_{\sigma,n}^*]
=2i\sigma\delta_{nm}\delta_{\sigma\sigma'},
\end{equation}
where $\sigma=1$ for outgoing states and  $\sigma=-1$ for incoming
states. Two popular choices are states with fixed angular momentum in
free space, and transversely quantized modes in a fixed-width waveguide
geometry.

We now describe the adaptation of this formalism to $\mathcal{PT}$ and
$\mathcal{PTT}'$-symmetric situations
\cite{hsqoptics,longhilaserabsorber,variousstonepapers1,
variousstonepapers2,yoo,hsrmt,birchallweyl}.

\subsection{${\cal PT}$ symmetry }

The symmetries of a scattering problem are exposed when one conveniently
groups the scattering states. To inspect ${\cal P}$ we call half of the
incoming states `incoming from the left', and the other half `incoming
from the right'. This does not need to be taken literally; all what
matters is that the two groups are converted into each other by ${\cal
P}$. The same can be done for the outgoing states. The scattering matrix
then  decomposes into blocks,
\begin{equation}
S=\left(
  \begin{array}{cc}
    r & t' \\
    t & r' \\
  \end{array}
\right),
\end{equation}
where $r$ describes reflection of left incoming states into left outgoing
states, $r'$ describes the analogous reflection on the right, while $t$
and $t'$  describe transmission from the left to the right and vice
versa. The parity $\mathcal{P}$ interchanges the amplitudes of the left
and right states, which can be brought about by a $\sigma_x$ Pauli
matrix,
\begin{equation}
\mathcal{P}[S]=\sigma_x S \sigma_x=
\left(\begin{array}{cc}
    r' & t \\
    t' & r \\
  \end{array}
\right).
\end{equation}

For consideration of the $\mathcal{T}$ operation, we group incoming and
outgoing states into time-reversed pairs. When $\mathcal{T}$ acts on a
wave function the amplitudes become conjugated, while the frequency in
the wave equation changes to $\omega^*$. Furthermore, incoming states are
converted into outgoing states, so that the relation \eqref{eq:sdef} must
be inverted. Thus,
\begin{equation}
\mathcal{T}[S(\omega)]=\{S^{-1}(\omega^*)\}^*.
\end{equation}

In combination, we have \cite{hsqoptics,variousstonepapers1,yoo}
\begin{equation}\label{eq:pts}
\mathcal{PT}[S(\omega)]=\sigma_x\{S^{-1}(\omega^*)\}^*\sigma_x.
\end{equation}
A system with $\mathcal{PT}$ symmetry  is then characterized by the
invariance
\begin{equation}
\sigma_x\{S^{-1}(\omega^*)\}^*\sigma_x=S(\omega),
\end{equation}
which results in the constraint $\sigma_xS^*(\omega^*)\sigma_xS(\omega)=\openone$, or
\begin{align}\label{eq:ptcond}
t^{\prime*}(\omega^*)r(\omega)+r^*(\omega^*)t(\omega)&=0,\nonumber \\  
r^{\prime*}(\omega^*)t'(\omega)+t^*(\omega^*)r'(\omega)&=0,\nonumber \\
r^*(\omega^*)r'(\omega)+t^{\prime*}(\omega^*)t'(\omega)&=\openone,\nonumber \\  
r^{\prime*}(\omega^*)r(\omega)+t^{*}(\omega^*)t(\omega)&=\openone.
\end{align}

\subsection{${\cal PTT}'$ symmetry }
In hermitian problems the scattering matrix is unitary if $\omega$ is
real, and the ${\cal T}$ operation is equivalent to the operation
\cite{beenakkerreview,datta}
\begin{equation}
\mathcal{T}'[S(\omega)]=S^T(\omega),
\end{equation}
which corresponds to the operation identified above by inspection of the
wave equation. In non-hermitian settings, this delivers the solution of
the scattering problem associated with the transposed wave equation
$\mathcal{L}^T\bar\psi(\mathbf{r})=0$,
\begin{equation}
\label{eq:st}
\bar{\mathbf{a}}_{\mathrm{out}}=S^T(\omega)\bar{\mathbf{a}}_{\mathrm{in}}.
\end{equation}

In ordinary optics, where $\textbf{A}=0$, $\mathcal{T}'$ remains an exact
symmetry even when $n$ and $\omega$ are complex. We then find the
important constraint $S(\omega)=S^T(\omega)$, thus
\begin{equation}\label{eq:sst}
r=r^T,\quad r'=r'^T,\quad t'=t^T.
\end{equation}
The combined operation
\begin{equation}\label{eq:ptts}
\mathcal{PTT}'[S(\omega)]=\sigma_x\{S^{-1}(\omega^*)\}^\dagger\sigma_x
\end{equation}
turns into a symmetry if \cite{hsrmt}
\begin{align}\label{eq:pttcond}
t^{\dagger}(\omega^*)r(\omega)+r^\dagger(\omega^*)t(\omega)&=0,\nonumber \\ 
r^{\prime\dagger}(\omega^*)t'(\omega)+t^{\prime\dagger}(\omega^*)r'(\omega)&=0,
\nonumber \\
r^\dagger(\omega^*)r'(\omega)+t^{\dagger}(\omega^*)t'(\omega)&=\openone,\nonumber \\ 
r^{\prime\dagger}(\omega^*)r(\omega)+t^{\prime\dagger}(\omega^*)t(\omega)&=\openone.
\end{align}
This is realized for systems obeying the symmetry requirements
\eqref{eq:ptna}.

\section{\label{sec:6}Generalized flux conservation}
For hermitian systems ($\omega$, $n$ and $\mathbf{A}$ all real), the
unitarity $S^\dagger S=\openone$ of the scattering matrix ensures the
conservation of the probability flux,  $\mathbf{a}_{\mathrm{in}}^\dagger
\mathbf{a}_{\mathrm{in}}=\mathbf{a}_{\mathrm{out}}^\dagger
\mathbf{a}_{\mathrm{out}}$. In Ref. \cite{variousstonepapers2} analogous
conservation laws were  established for  one-dimensional non-hermitian
systems with $\mathcal{PT}$ symmetry, and it was found that these laws
are automatically guaranteed by the symmetry conditions
\eqref{eq:ptcond}. These flux conditions only apply at real $\omega$, but
provide a useful alternative perspective on the role of symmetry. Here we
extend these conditions to higher-dimensional (multichannel) systems,
include magneto-optical effects, and  also allow for $\mathcal{PTT}'$
symmetry (these considerations are original).

To illustrate the general strategy we consider the Helmholtz equation for
TM polarization, with ${\bf A}=0$ and $\omega$ real. Thus
$\psi(\mathbf{r})$ and $\psi^*(\mathcal{P}\mathbf{r})$ both solve the
same wave equation $\mathcal{L}(\omega)\psi(\mathbf{r})
=\mathcal{L}(\omega)\psi^*(\mathcal{P}\mathbf{r})=0$, with
$\mathcal{L}(\omega)$ specified in equation \eqref{eq:helmholtz}. Now take the
following volume integral over the region $\Omega$,
\begin{align}
0&=\int_\Omega d\mathbf{r}\,
[\psi^*(\mathcal{P}\mathbf{r})\mathcal{L}(\omega)\psi(\mathbf{r})
-\psi(\mathbf{r})\mathcal{L}(\omega)\psi^*(\mathcal{P}\mathbf{r})]
\nonumber \\  &=
\int_\Omega d\mathbf{r}\,
[\psi^*(\mathcal{P}\mathbf{r})\nabla^2 \psi(\mathbf{r})
-\psi(\mathbf{r})\nabla^2 \psi^*(\mathcal{P}\mathbf{r})]
\nonumber \\  &=
\int_\Omega d\mathbf{r}\,\nabla
[\psi^*(\mathcal{P}\mathbf{r})\nabla \psi(\mathbf{r})
-\psi(\mathbf{r})\nabla \psi^*(\mathcal{P}\mathbf{r})]
\nonumber \\  &=
\int_{\partial\Omega}d\mathbf{S}\cdot
[\psi^*(\mathcal{P}\mathbf{r})\nabla \psi(\mathbf{r})
-\psi(\mathbf{r})\nabla \psi^*(\mathcal{P}\mathbf{r})],
\end{align}
under application of Stoke's theorem. The resulting surface integral can
be evaluated with equation \eqref{eq:flux}. This delivers the generalized
flux-conservation law
\begin{equation}
\mathbf{a}_{\mathrm{in}}^\dagger\sigma_x \mathbf{a}_{\mathrm{in}}=\mathbf{a}_{\mathrm{out}}^\dagger\sigma_x \mathbf{a}_{\mathrm{out}}=\mathbf{a}_{\mathrm{in}}^\dagger S^\dagger(\omega)\sigma_x S(\omega)\mathbf{a}_{\mathrm{in}},
\end{equation}
which together with $S(\omega)=S^T(\omega)$ (as $\mathbf{A}=0$) amounts
to the constraints \eqref{eq:ptcond}, specialized to the case where
$\omega$ is real.

The same constraints follow for TE polarization (as $n=1$ outside
$\Omega$, so that the refractive index does not feature in the surface
integral). Next, we include a finite vector potential $\textbf{A}$. For
$\mathcal{PTT}'$ symmetry, one can follow the steps given above, which
directly delivers the constraints \eqref{eq:pttcond} (again specialized
to real $\omega$). The case of $\mathcal{PT}$ symmetry with finite vector
potential is more intricate, as one then needs to invoke  the transposed
wave equation
$\mathcal{L}^T(\omega)\bar\psi(\mathbf{r})=\mathcal{L}^T(\omega)\bar\psi^*(\mathcal{P}\mathbf{r})=0$.
One then integrates
\begin{align}
0&=\int_\Omega d\mathbf{r}\,
[\bar\psi^*(\mathcal{P}\mathbf{r})\mathcal{L}(\omega)\psi(\mathbf{r})
-\psi(\mathbf{r})\mathcal{L}^T(\omega)\bar\psi^*(\mathcal{P}\mathbf{r})]
\nonumber \\  &=
\int_{\partial\Omega}d\mathbf{S}\cdot
[\bar\psi^*(\mathcal{P}\mathbf{r})\nabla \psi(\mathbf{r})
-\psi(\mathbf{r})\nabla \bar\psi^*(\mathcal{P}\mathbf{r})],
\end{align}
which results in the generalized flux-conservation  law
$\bar{\mathbf{a}}_{\mathrm{in}}^\dagger \sigma_x \mathbf{a}_{\mathrm{in}}
=\bar{\mathbf{a}}_{\mathrm{out}}^\dagger\sigma_x \mathbf{a}_{\mathrm{out}}
$.
In combination with equation \eqref{eq:st}, this condition now requires
$\sigma_x=S^*(\omega)\sigma_x S(\omega)$, which is again automatically
fulfilled if the scattering matrix obeys the symmetry constraints
\eqref{eq:ptcond}.

In all these cases, the generalized flux-conservation relations at real
$\omega$ thus does not impose any extra conditions beyond the symmetry
requirements of the scattering matrix.

\section{\label{sec:7}Scattering quantization conditions}
It is now interesting to ask how the spectral properties of
$\mathcal{PT}$ or $\mathcal{PTT}'$-symmetric systems emerge within the
scattering approach. We contrast the case of open systems (where the
symmetry operation relates quasibound states with perfectly absorbed
states, thus, connects states of a different physical nature) with closed
systems (where the symmetry operation relates ordinary bound states,
thus, connects states of the same nature, whose spectral properties then
are  constrained).

\subsection{General quantization conditions for quasibound and perfectly absorbed states}
Quasibound states fulfill the wave equation with purely outgoing boundary
conditions, thus, $\mathbf{a}_{\mathrm{in}}=0$ but
$\mathbf{a}_{\mathrm{out}}$ finite. In view of equation \eqref{eq:sdef},
this requires
\begin{equation}\label{eq:quant}
||S(\omega)||=\infty,
\end{equation}
and so $\omega$ has to coincide with a pole of the scattering matrix.
This results in a quantization of the admitted frequencies $\omega_n$,
which in general are complex.

Quasibound states describe systems which generate all radiation
internally, and in particular, lasers
\cite{hsqoptics,longhilaserabsorber,variousstonepapers1,variousstonepapers2,yoo}.
In a passive system the poles are confined to the lower half of the
complex plane, with the decay enforced by the leakage. In an active
system, however, gain may compensate the losses and result in a
stationary state, which signifies lasing. The threshold is attained when
the first pole reaches the real axis.

Recent works have turned the attention to perfectly absorbed states, for
which the role of incoming and outgoing states is inverted
\cite{longhilaserabsorber,stonecpa1}. These boundary conditions translate
to
$||S(\omega)||=0$,
which results in quantized frequencies  $\widetilde{\omega}_n$ coinciding
with the zeros of the scattering matrix.

For a passive system, where
$\mathcal{TT}'[\mathcal{L}(\omega)]=\mathcal{L}(\omega^*)$, a quasibound
state $\psi_n(\mathbf{r})$ can be converted into a perfectly absorbed
state $\widetilde\psi_n(\mathbf{r})=\bar\psi_n^*(\mathbf{r})$ by the
$\mathcal{TT}'$ operation [for exact $\mathcal{T}'$ symmetry one simply
has $\widetilde\psi_n(\mathbf{r})=\psi_n^*(\mathbf{r})$]. This state then
fulfills the  wave equation at $\widetilde{\omega}_n=\omega_n^*$, which
confines the zeros to the upper half of the complex plane.

For non-hermitian systems the relation between the poles and zeros is in
general broken. With $\mathcal{PT}$ or $\mathcal{PTT}'$ invariance,
however, the states $\widetilde\psi_n(\mathbf{r})=\psi_n^*({\cal
P}\mathbf{r})$ or $\widetilde\psi_n(\mathbf{r})=\bar\psi^*_n({\cal
P}\mathbf{r})$ are paired by the respective symmetry. Moreover, the
frequencies $\omega_n=\widetilde{\omega}_n^*$  then are no longer
constraint to the lower half of the complex plane. At the lasing
condition $\omega_n=0$, the lasing mode is then degenerate with a
perfectly absorbed mode, which leads to the concept of a
$\mathcal{PT}$-symmetric laser-absorber \cite{longhilaserabsorber}.

\subsection{Closed resonators and spectral constraints}
By taking the appropriate limit of the expressions for open systems, the
scattering approach to mode quantization can be used to study closed
systems, for which  $\mathcal{PT}$ symmetry was originally defined. The
quasibound states then turn into normal bound states, and one recovers
the spectral constraints \eqref{eq:constraint}. These considerations also
apply to $\mathcal{PTT}'$ symmetry. They can also be based on the
perfectly absorbed states, which become degenerate with the quasibound
states.

When one reduces the leakage from a non-hermitian $\mathcal{PT}$ or
$\mathcal{PTT}'$-symmetric optical system, it will ultimately start to
lase. If some frequencies of the closed system are complex then lasing is
attained at a finite leakage
\cite{longhilaserabsorber,variousstonepapers1,variousstonepapers2,yoo};
otherwise one approaches at-threshold lasing \cite{hsqoptics}.

\section{\label{sec:8}
Coupled-resonator geometry}

\begin{figure}
\includegraphics[width=\columnwidth]{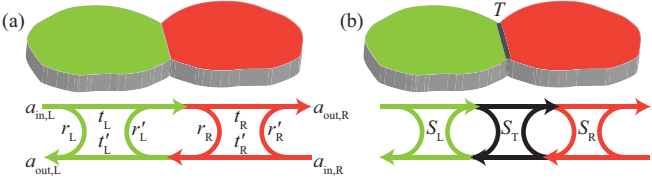}
\caption{\label{fig3}(Online version in colour.)
Scattering amplitudes and their relation by reflection and transmission
blocks of the scattering matrix, for the coupled-resonator geometry
considered in \S\ref{sec:8}--\S\ref{sec:10}. In (b), a finite
transparency $T$ of the interface is taken into account. See also
\cite{hsqoptics,yoo,hsrmt}. }
\end{figure}

The economy of the scattering approach is full exposed when one applies
it to multichannel systems capable of displaying complex wave dynamics.
Here we review this for a broad class of systems in which an absorbing
resonator is coupled symmetrically to an amplifying resonator, as shown
in figure \ref{fig3}  \cite{hsqoptics,yoo,hsrmt,birchallweyl}.

\subsection{Scattering matrix}

Denoting the scattering matrices of the two resonators as
\begin{equation}
S_L=\left(
      \begin{array}{cc}
        r_L & t_L' \\
        t_L & r_L' \\
      \end{array}
    \right), \quad S_R=\left(
      \begin{array}{cc}
        r_R & t_R' \\
        t_R & r_R' \\
      \end{array}
    \right),
\end{equation}
the scattering matrix $S=S_L\circ S_R$ of the composed system is given by
\begin{align}
&\left( \begin{array}{cc}r_L&t_L'\\t_L&r_L'\end{array} \right)\circ
\left( \begin{array}{cc}r_R&t_R'\\t_R&r_R'\end{array} \right)
\nonumber \\ &=
\left( \begin{array}{cc}r_L+t_L'\frac{1}{1-r_Rr_L'}r_Rt_L&t_L'\frac{1}{1-r_Rr_L'}t_R'
\\ t_R\frac{1}{1-r_L'r_R}t_L&r_R'+t_R\frac{1}{1-r_L'r_R}r_L't_R'\end{array} \right).
\label{eq:comprule}
\end{align}
$\mathcal{PT}$ symmetry follows if the scattering matrices are related by
$S_R=\mathcal{PT}[S_L]$, with this operation specified in equation
\eqref{eq:pts}, while $\mathcal{PTT}'$ symmetry is realized if the
scattering matrices are related by $S_R=\mathcal{PTT'}[S_L]$, as
specified in equation \eqref{eq:ptts}.

This construction can be amended to include an interface of finite
transparency $T$, with the scattering matrix, \emph{e.g.}, specified by
\begin{equation}\label{eq:stun}
S_T=
\left(\begin{array}{cc}-\sqrt{1-T} & i\sqrt{T}
\\ i\sqrt{T} & -\sqrt{1-T} \\ \end{array}\right).
\end{equation}
The total scattering matrix of the system is then given by $S=S_L\circ
S_T\circ S_R$.

\subsection{Quantization conditions and spectral constraints}

With equation \eqref{eq:comprule} the scattering quantization condition
\eqref{eq:quant} becomes
\begin{equation}
{\rm det}\,[r_L'(\omega) r_R(\omega) -\openone]=0.
\end{equation}
For a $\mathcal{PT}$-symmetric system obeying equation \eqref{eq:ptcond}
this condition takes the form
\begin{equation}\label{eq:ptquant}
{\rm det}\,(r_L'(\omega)-
[r_L'(\omega^*)-t_L(\omega^*)r_L^{-1}(\omega^*)t_L'(\omega^*)]^*)=0,
\end{equation}
while for a $\mathcal{PTT}$-symmetric system obeying equation
\eqref{eq:pttcond} this gives
\begin{equation}\label{eq:ptquantptt}
{\rm det}\,(r_L'(\omega)-
[r_L'(\omega^*)-t_L(\omega^*)r_L^{-1}(\omega^*)t_L'(\omega^*)]^\dagger)=0.
\end{equation}

For a closed system the scattering matrices reduce to the reflection
blocks $S_L(\omega)=r_L'(\omega)$ and $S_R(\omega)=r_R(\omega)$, while
the transmission vanishes. In the case of $\mathcal{PT}$ symmetry, with
$r_R(\omega)= \{[r_L'(\omega^*)]^{-1}\}^*$, equation \eqref{eq:ptquant}
assumes the form
\begin{equation}\label{eq:ptquant2}
{\rm det}\,(r_L'(\omega)-[r_L'(\omega^*)]^*)=0,
\end{equation}
which entails the constraints \eqref{eq:constraint}. On the real
frequency axis, this reduces to the condition $ {\rm det}\,{\rm
Im}\,r_L'(\omega)=0. $ Analogous observations also hold for closed
$\mathcal{PTT}'$-symmetric resonators, for which the quantization
condition  \eqref{eq:ptquantptt} becomes
\begin{equation}\label{eq:ptquant2ptt}
{\rm det}\,(r_L'(\omega)-[r_L'(\omega^*)]^\dagger)=0.
\end{equation}

Under inclusion of a semitransparent interface, similar conditions can be
derived from the general expression
\begin{equation}\label{eq:ptquantst} {\rm
det}\,\left[S_T\left(\begin{array}{cc}r_L' & 0 \\0 & r_R \\
\end{array}\right)-\openone\right]=0.
\end{equation}

These scattering quantization conditions can all be interpreted as
conditions for constructive interference upon return to the interface
between the two parts of the resonator. The analogous conditions for
perfectly absorbed states follow from the replacement $\omega\to
\omega^*$.

\section{\label{sec:9}Effective Models}
The coupled-resonator geometry allows to make contact to well-studied
standard descriptions of multiple scattering
\cite{beenakkerreview,fyodorovsommers,fyodsomeffs,andreev,schomerusjacquod}.
This leads to effective models of complex wave dynamics in systems with
$\mathcal{PT}$ and $\mathcal{PTT}'$ symmetry, which we first formulate in
the energy domain \cite{hsrmt}, and then in the time domain
\cite{birchallweyl}.

\subsection{Effective Hamiltonians}
We first employ the Hamiltonian approach to multiple scattering
\cite{beenakkerreview,fyodorovsommers}. The hermitian part of the
dynamics in the absorbing resonator is captured by a hermitian  $M\times
M$-dimensional matrix $H$. We assume $M\gg 1$ and denote the level
spacing in the energy range of interest as $\Delta$. Loss with absorption
rate $\mu$ is modeled by adding a non-hermitian term $-i\mu\openone$,
while gain with a matching rate is obtained by inverting the sign of
$\mu$. We also specify an $N\times M$-dimensional  coupling matrix $V$
between the $M$ internal modes and $N$ scattering states. This matrix can
be chosen to satisfy $V^TV={\rm diag}{(v_m)}$, where  $N$ finite entries
$v_m=\Delta M/\pi$ describe the open channels, while $M-N$ entries
$v_m=0$ describe the closed channels. The $N\times N$-dimensional
scattering matrix of the absorbing resonator is then given by
\begin{equation}
S_L(\omega) = 1-2i V (\omega-i\mu-H+i V^TV)^{-1}V^T.
\end{equation}
Thus, the leakage from the system effectively adds an additional
non-hermitian term $-i V^TV$ to the Hamiltonian. For $\mathcal{PT}$
symmetry the scattering matrix of the amplifying resonator follows as
\cite{hsrmt}
\begin{equation}
S_R(\omega)=[S_L^{-1}(\omega^*)]^*= 1-2i V (\omega+i\mu-H^*+i V^TV)^{-1}V^T.
\end{equation}
We also include a semitransparent interface, with scattering matrix
\eqref{eq:stun}. For a closed system, the scattering quantization
condition \eqref{eq:ptquantst} can then be rearranged into an eigenvalue
problem ${\rm det}\,(\omega-{\cal H})=0$ with effective Hamiltonian
\cite{hsrmt}
\begin{equation}
\label{eq:h}
{\cal H}=\left(\begin{array}{cc}H-i\mu &   \Gamma  \\   \Gamma & H^*+i\mu \\ \end{array}\right).
\end{equation}
Here $\Gamma={\rm diag}\,(\gamma_m)$ is a real positive semi-definite
coupling matrix related to $V^TV$, but with the $N$ non-vanishing entries
$\gamma_m=[\sqrt{T}/(1+\sqrt{1-T})]\Delta M/\pi\equiv \gamma$ modified to
account for the finite transparency $T$ of the interface.

The effective Hamiltonian obeys the relation
$\mathcal{PT}[\mathcal{H}]=\sigma_x \mathcal{H}^* \sigma_x =\mathcal{H}$.
In a $\mathcal{PT}$-symmetric basis the secular equation ${\rm
det}\,(\omega-{\cal H})=0$ then takes the form of a polynomial with real
coefficients, which guarantees that the eigenvalues are either real or
occur in complex-conjugate pairs, as required by equation
\eqref{eq:constraint}.

Analogous considerations apply to $\mathcal{PTT}'$-symmetric systems
\cite{hsrmt,birchallrmt}. The scattering matrix of the right resonator
then changes to
\begin{equation}
S_R(\omega)=[S_L^{-1}(\omega^*)]^\dagger= 1-2i V (\omega+i\mu-H+i V^TV)^{-1}V^T,
\end{equation}
and the effective Hamiltonian of the closed system takes the form
\begin{equation}
\label{eq:hptt}
{\cal H}=\left(\begin{array}{cc}H-i\mu &   \Gamma  \\   \Gamma & H+i\mu \\ \end{array}\right).
\end{equation}
We then have
$\mathcal{PTT}'[\mathcal{H}]=\sigma_x \mathcal{H}^\dagger \sigma_x =\mathcal{H}$,
which again guarantees the required spectral constraints.

While Hamiltonians with these symmetries could simply be stipulated, the
derivation of the specific manifestations \eqref{eq:h} and
\eqref{eq:hptt} reveals further constraints dictated by the
coupled-resonator geometry. In particular, the coupling is only physical
if the matrix $\Gamma$ is positive semidefinite. These Hamiltonians bear
a striking resemblance to models of mesoscopic superconductivity
\cite{beenakkerreview,melsen}.


\subsection{Quantum maps}

\begin{figure}
\includegraphics[width=\columnwidth]{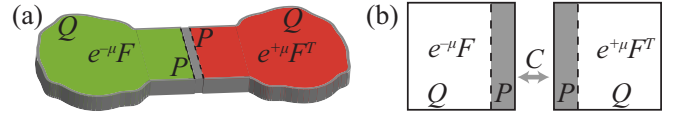}
\caption{\label{fig4}(Online version in colour.)
Interpretation of the $\mathcal{PT}$-symmetric quantum map
\eqref{eq:map}, which translates the resonator dynamics in (a) to a
stroboscopic evolution in two coupled Hilbert spaces (b). For the
$\mathcal{PTT}'$-symmetric map \eqref{eq:mapptt}, $F^T$ is replaced by
$F$. See also \cite{birchallweyl}. }
\end{figure}

Alternatively, one can set up effective descriptions in the time domain
\cite{birchallweyl}. For $\mathcal{PT}$ symmetry this sets out by writing
the scattering matrices as \cite{fyodsomeffs,andreev,schomerusjacquod}
\begin{subequations}
\begin{align}
S_L(\omega)&=WF[\exp(-i\omega\tau+\mu)-QF]^{-1}W^T,\\
S_R(\omega)&=WF^T[\exp(-i\omega\tau-\mu)-QF^T]^{-1}W^T,
 \label{eq:smat}
\end{align}
\end{subequations}
where $F$ is an  $M\times M$-dimensional unitary matrix which can be
thought to describe the stroboscopic time evolution between successive
scattering events off the resonator walls. The transfer across the
interface between the two resonators is again described by an $N\times
M$-dimensional coupling matrix $W$. However, the combination $P=W^TW$ now
projects the wave function onto the interface, such that ${\rm
rank}\,P=N$ is the number of channels connecting the resonators, while
$Q=\openone_M-P$ is the complementary projector onto the resonator wall
(${\rm rank}\,Q=M-N$). The stated frequency dependence corresponds to
stroboscopic scattering with fixed rate $\tau^{-1}$; the variable
$\omega$ thus plays the role of a quasienergy. The parameter $\mu\geq 0$
again determines the absorption and amplification rate. For the passive
system ($\mu=0$), the scattering matrices are unitary, which corresponds
to the hermitian limit of the problem.

With these specifications, and including a semitransparent interface
parameterized by $\alpha\equiv\sqrt{\sqrt{R}+i\sqrt{T}}$, the
quantization condition (\ref{eq:ptquantst}) is equivalent to the
eigenvalue problem
\begin{equation}
{\cal F}\psi_n=\lambda_n\psi_n,\quad \lambda_n=\exp(-i\omega_n\tau)
\end{equation}
for the quantum map
\begin{align}\label{eq:map}
{\cal F}&=
\sqrt{C}
\left(\begin{array}{cc} e^{-\mu}F & 0 \\ 0 &  e^{\mu}F^T \\ \end{array}  \right)
\sqrt{C},
\nonumber \\
\quad \sqrt{C}&=
\left(\begin{array}{cc} \mathrm{Re}\,\alpha\,P+Q & -i\mathrm{Im}\,\alpha\,P\\
-i\mathrm{Im}\,\alpha\,P  &\mathrm{Re}\,\alpha\,P+Q \\ \end{array}  \right)
\end{align}
where we symmetrized the coupling
\begin{equation}
C=\sqrt{C}\sqrt{C}=\left(\begin{array}{cc} \sqrt{R}P+Q &
-iP\sqrt{T} \\ -iP\sqrt{T} & \sqrt{R}P+Q \\ \end{array} \right).
\end{equation}
The $2M\times 2M$ dimensional matrix ${\cal F}$ can be interpreted as a
stroboscopic time evolution operator acting on $2M$-dimensional vectors
$\psi=\binom{\psi_L}{\psi_R}$, where $\psi_L$ and $\psi_R$ give the wave
amplitude in the absorbing and amplifying subsystem, respectively. This
is illustrated in figure \ref{fig4}. The $\mathcal{PT}$ symmetry of the
quantum map manifests itself in the relation
${\cal F}=\sigma_x[{\cal F}^{-1}]^*\sigma_x$,
which parallels the symmetry \eqref{eq:pts} for the scattering matrix.

Analogously, $\mathcal{PTT}'$ symmetry results in the quantum map
\begin{equation}
{\cal F}=
\sqrt{C}
\left(\begin{array}{cc} e^{-\mu}F & 0 \\ 0 &  e^{\mu}F \\ \end{array}  \right)
\sqrt{C}.
\label{eq:mapptt}
\end{equation}
This obeys the symmetry
${\cal F}=\sigma_x[{\cal F}^{-1}]^\dagger\sigma_x$,
which parallels equation \eqref{eq:ptts}.

The spectral properties associated with these symmetries are now embodied
in the secular equation $s(\lambda)=\rm{det}\,({\cal F} -\lambda)=0$,
which exhibits the mathematical property of \emph{self-inversiveness}
\cite{selfinverse2}:
\begin{equation}
s(1/\lambda^*)=[\lambda^{-2M}s(\lambda)]^*s(0),
\end{equation}
where $s(0)=\mathrm{det}\,\mathcal{F}= (\mathrm{det}\,F)^2$. For each
eigenvalue $\lambda_n$, we are thus guaranteed to find the eigenvalue
$\lambda_{\bar n}=[\lambda_n^{-1}]^*=\exp(-i\omega_n^*\tau)$, which
recovers the constraint \eqref{eq:constraint}.

\section{\label{sec:10}Mesoscopic energy and times scales}

In mesoscopic physics, a large range of spectral, thermodynamic and
transport  phenomena are fully characterized by a few universal time and
energy scales \cite{beenakkerreview,datta,altlandthouless}. We now have
all the tools to make contact to these concepts and provide a
phenomenological description of the effects of multiple scattering on the
spectral features  of the considered coupled resonators. These effects
generally set in when a large number $M\gg1$ of internal modes in an
energy range $M\Delta$ is mixed by scattering with a characteristic time
scale $\tau=\hbar/M\Delta$ that is much less than the dwell time in the
amplifying or absorbing regions, $\tau\ll t_{\mathrm{dwell}}$. The
coupling strength between these regions is characterized by the
associated Thouless energy $E_T=\hbar/t_{\mathrm{dwell}}\approx NT
\Delta$, which can be small or large compared to the level spacing
$\Delta$, depending on the value of the dimensionless conductance $g=NT$
of the interface. Non-hermiticity adds the new scale $\mu$, whose
interplay with the other scales determines the transition from real to
complex eigenvalues.

\begin{figure}
\includegraphics[width=\columnwidth]{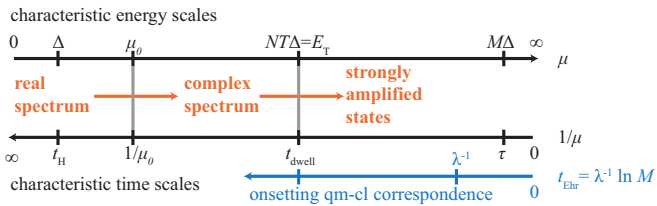}
\caption{\label{fig5}(Online version in colour.)
Sketch of the energy and time scales governing the spectral features of
$\mathcal{PT}$-symmetric resonators with separate
$\mathcal{T}'$ invariance (no magneto-optical effects), as hermiticity is
broken with absorption and amplification rate $\mu$. This universal
picture holds under the assumption of complex wave dynamics ($M\gg
N,NT\gg 1$), where $M=\hbar/\tau\Delta$ is the number of internal modes
mixed by multiple scattering with scattering time $\tau$ and $N$ is the
number of channels coupling the resonators, with transparency $T$. Here
$\Delta$ is the mean level spacing,  $t_{\rm H}=\hbar/\Delta$ the
Heisenberg time, and $E_T=\hbar/t_{\rm dwell}=NT\Delta$ the Thouless
energy associated with the dwell time $t_{\rm dwell}$ in each resonator.
The critical rate at which eigenvalues turn complex is
$\mu_0=\sqrt{N}\Delta/2\pi$. When the Ehrenfest time $t_{\rm
Ehr}=\lambda^{-1}\ln M$ (with Lyapunov exponent $\lambda$) becomes
comparable to the dwell time $t_{\rm dwell}$  quantum-to
classical-correspondence sets in and suppresses multiple scattering, which reduces the
number of strongly amplified states. See also \cite{hsrmt,birchallweyl}.
}
\end{figure}

In the mesoscopic regime of $M\gg N \gg 1$ this transition can be
expected to be universal, with the features only depending on $M/N \gg
1$, $M\gg 1$, and $T$, which we consider fixed by the geometry, as well
as the variable  $\mu$. The transition can then be investigated by
combining the effective models set up in the previous section with
random-matrix theory \cite{beenakkerreview,haake,mehta}, thus, ensembles
of Hamiltonians $H$ (usually composed with random Gaussian matrix
elements) or time-evolution operators $F$ (distributed according to a
Haar measure) which are only constrained by the symmetries of the
problem. Closer inspection identifies two natural scenarios
\cite{hsrmt,birchallrmt}, described in the following two subsections,
and a semiclassical source of corrections to random-matrix theory, discussed thereafter \cite{birchallweyl}. The interplay of the various time and energy scales is illustrated in
figure \ref{fig5}.

\subsection{Systems with coupling-driving level crossings}

In ordinary optics with exact $\mathcal{T}'$ symmetry, systems with
$\mathcal{PT}$ symmetry also display $\mathcal{PTT}'$ symmetry. The
internal Hamiltonian $H=H^*=H^T$ is then real and symmetric, and the
effective Hamiltonians \eqref{eq:h} and  \eqref{eq:hptt} coincide.
(Analogously, $F=F^T$, so that the quantum maps \eqref{eq:map} and
\eqref{eq:mapptt} coincide as well.) Switching to a parity-invariant
basis one then finds
\begin{equation}\label{eq:hpar}
{\cal H}_{\cal P}=\left(\begin{array}{cc}H+\Gamma &   i\mu  \\  i\mu  & H-\Gamma \\ \end{array}\right),
\end{equation}
which reveals the emerging $\mathcal{T}$ symmetry in the hermitian limit.
The same structure arises for systems with magneto-optical effects which
only display  $\mathcal{PTT}'$. For both cases, at $\mu=0$ the system
decouples into two independent sectors. For $T=0$ the level sequences of
these sectors coincide. For very weak coupling ($T\ll 1/N\equiv T_c$,
thus $g\ll 1$) the degeneracy is slightly lifted, and one can apply
almost-degenerate perturbation theory to identify the scale $\mu\approx
N\sqrt{T}\Delta/2\pi$ at which typical eigenvalues turn complex. As $T$
exceeds $1/N\equiv T_c$ ($g>1$), however, the levels of the original
sequences cross, and when one then increases $\mu$ the bifurcations to
complex eigenvalues occur between levels that were originally
non-degenerate. In this regime, a macroscopic fraction of complex
eigenvalues appears on a \emph{coupling-independent} scale
$\mu\approx\mu_0\equiv\sqrt{N}\Delta/2\pi$.

\subsection{Systems with coupling-driving avoided crossings}
For systems  with $\mathcal{PT}$ symmetry but broken $\mathcal{T}'$
symmetry, the parity basis does not partially diagonalize the effective
Hamiltonian. At $\mu=T=0$, one still starts from two degenerate level
sequences, but these interact as $T$ is increased, and instead of level
crossings one observes level repulsion. In this case the crossover to a
complex spectrum appears at $\mu\approx \sqrt{NT}\Delta/2\pi$ and thus is
always coupling dependent.

\subsection{Growth and decay rates in the semiclassical limit}

In the semiclassical limit of large $M$ at fixed $M/N$ the results above
imply that the crossover to a complex spectrum appears at a vanishingly
small rate $\mu\ll E_T$. Keeping $\mu/E_T$ in this limit fixed and
finite, the real phase is thus completely destroyed, and the focus turns
to the typical decay and growth rates encoded in the imaginary parts
${\rm Im}\, \omega_n$ of the complex eigenvalues. Numerical sampling of
the random-matrix ensembles suggests that the distribution $P({\rm Im}\,
\omega_n/\mu;\mu/E_T)$ attains a stationary limit for $M\gg 1$
\cite{hsrmt,birchallweyl,birchallrmt}. At $\mu\gtrsim E_T$, one then
finds a finite fraction of strongly amplified states with  ${\rm Im}\,
\omega_n\approx \mu$, thus, a large number of candidate lasing states.
However, the fraction of these states is reduced when one takes dynamical
effects into account  \cite{birchallweyl}. One then finds that the
strongly amplified states are supported by the classical  repeller in the
amplifying parts of the system, whose fractal dimension $d_H$ is more and
more resolved as the phase space resolution $h\propto 1/M$ increases in
the semiclassical limit. This phenomenon can be characterized by the
Ehrenfest time $t_{\rm Ehr}=\lambda^{-1}\ln M$, where $\lambda$ the
Lyapunov exponent in the classical system. If $t_{\rm Ehr}>t_{\rm
dwell}$, the wave dynamics become quasi-deterministic, and multiple
scattering is reduced. The fraction of strongly amplified states then
follows a fractal Weyl law (a power law in $h$ with non-integer
exponent), a phenomenon which was previously  observed for passive
quantum systems with finite leakage through ballistic openings
 \cite{fractalweyl1,fractalweyl3}.

\section{\label{sec:11}Summary and outlook}

In summary, the scattering approach proves useful to describe general
features of non-hermitian optical systems with $\mathcal{PT}$ and
$\mathcal{PTT}'$ symmetry. In particular, the approach fully accounts for
complications that arise beyond one dimension (a choice of non-equivalent
geometric and time-reversal symmetries, the possibility of
magneto-optical effects with a finite external vector potential, and
multiple scattering), as well as additional leakage which turns the
symmetry of bound states into a relation between quasibound and perfectly
absorbed states. These effects can be captured in effective model
Hamiltonians and time evolution operators, which help to identify the
mesoscopic energy and time scales that govern the spectral features of a
broad class of systems. The construction of these models exposes physical
constraints that go beyond the mere symmetry requirements.

The models formulated here can be adapted to include, \emph{e.g.},
inhomogeneities in the gain, dissipative magneto-optical effects, leakage, or other symmetry-breaking effects \cite{birchallrmt}. Furthermore,
the construction of symmetric resonators from two subsystems can be
extended to include more elements, such as they occur in periodic or
disordered arrays. This bridges to models of higher-dimensional diffusive
or Anderson-localized dynamics \cite{west}. The models can also be
generalized to include symplectic time-reversal symmetries (with
$\mathcal{T}^2=-1$), chiral symmetries, or particle-hole
symmetries (as the $\mathcal{CT}$ symmetry obeyed by mesoscopic superconductors), for which the
consequences on the scattering matrix and effective Hamiltonians are
well known in the hermitian case \cite{beenakkerreview,altlandzirnbauer,melsen}.  Some of these
effects have optical analogues; \emph{e.g.}, phase-conjugating mirrors
induce effects similar to Andreev reflection in mesoscopic
superconductivity but result in a non-hermitian Hamiltonian \cite{beenakkerphase}. Finally we note that the scattering approach also
offers a wide range of analytical and numerical methods which allow to
efficiently study individual systems \cite{datta}.

I thank Christopher Birchall for fruitful collaboration on Refs.\
\cite{birchallweyl,birchallrmt}, which form the basis of some of the
material reviewed here, as well as Uwe G{\"u}nther for fruitful
discussions concerning the role of symmetries.


\begin{thebibliography}{98}

\bibitem{bender}
Bender, C. M. \& Boettcher, S. 1998
Real Spectra in Non-Hermitian Hamiltonians Having PT Symmetry
\emph{Phys. Rev. Lett.} \textbf{80}, 5243--5246.
(doi:10.1103/PhysRevLett.80.5243)

\bibitem{ptreview1}
Bender, C. M. 2007
Making sense of non-Hermitian Hamiltonians.
\emph{Rep. Prog. Phys.} \textbf{70}, 947--1018.
(doi:10.1088/0034-4885/70/6/R03)

\bibitem{ptreview2}
Mostafazadeh, A. 2010
Pseudo-Hermitian representation of quantum mechanics.
\emph{Int. J. Geom. Meth. Mod. Phys.} \textbf{7}, 1191--1306.
(doi:10.1142/S0219887810004816)


\bibitem{christodoulidesgroup1}
El-Ganainy, R., Makris, K. G.,  Christodoulides, D. N. \&  Musslimani, Z. H. 2007
Theory of coupled optical PT-symmetric structures.
\emph{Opt. Lett.} \textbf{32}, 2632--2634. (doi:10.1364/OL.32.002632)

\bibitem{christodoulidesgroup2}
Makris, K. G.,
El-Ganainy, R., Christodoulides, D. N. \& Musslimani,  Z. H. 2008
Beam Dynamics in PT Symmetric Optical Lattices.
\emph{Phys. Rev. Lett.} \textbf{100}, 103904.
(doi:10.1103/PhysRevLett.100.103904)


\bibitem{christodoulidesgroup3}
Musslimani, Z. H., Makris, K. G., El-Ganainy, R. \& Christodoulides, D. N. 2008
Optical Solitons in PT Periodic Potentials.
\emph{Phys. Rev. Lett.} \textbf{100}, 030402.
(doi:10.1103/PhysRevLett.100.030402)

\bibitem{ptexperiments1}
Guo, A., Salamo, G. J., Duchesne, D., Morandotti, R., Volatier-Ravat, M., Aimez, V., Siviloglou, G. A. \&  Christodoulides, D. N. 2009
Observation of PT-Symmetry Breaking in Complex Optical Potentials.
\emph{Phys. Rev. Lett.} \textbf{103}, 093902.
(doi:10.1103/PhysRevLett.103.093902)

\bibitem{ptexperiments2}
R{\"u}ter, C. E., Makris, K. G., El-Ganainy,
R., Christodoulides, D. N., Segev, M. \& Kip, D. 2010
Observation of parity-time symmetry in optics.
\emph{Nature Phys.} \textbf{6}, 192--195.
(doi:10.1038/nphys1515)

\bibitem{pteffects1}
Longhi, S. 2009
Bloch Oscillations in Complex Crystals with PT Symmetry.
\emph{Phys. Rev. Lett.} \textbf{103}, 123601.
(doi:10.1103/PhysRevLett.103.123601)

\bibitem{pteffects2a}
Ramezani, H.,  Kottos, T., El-Ganainy, R. \&
Christodoulides, D. N. 2010
Unidirectional nonlinear PT-symmetric optical structures.
\emph{Phys. Rev. A} \textbf{82}, 043803.
(doi:10.1103/PhysRevA.82.043803)

\bibitem{pteffects4}
Longhi, S. 2011
Invisibility in PT-symmetric complex crystals.
\emph{J. Phys. A: Math. Theor.} \textbf{44}, 485302.
(doi:10.1088/1751-8113/44/48/485302)

\bibitem{pteffects5}
Lin, Z.,  Ramezani, H., Eichelkraut, T., Kottos, T., Cao, H. \&
Christodoulides, D. N. 2011
Unidirectional Invisibility Induced by PT-Symmetric Periodic Structures.
\emph{Phys. Rev. Lett.} \textbf{106}, 213901.
(doi:10.1103/PhysRevLett.106.213901)

\bibitem{hsqoptics}
Schomerus, H.  2010
Quantum Noise and Self-Sustained Radiation of PT-Symmetric Systems.
\emph{Phys. Rev. Lett.} \textbf{104}, 233601.
(doi:10.1103/PhysRevLett.104.233601)

\bibitem{longhilaserabsorber} Longhi, S.  2010 PT-symmetric laser
    absorber. \emph{Phys. Rev. A} \textbf{82}, 031801(R).
    (doi:10.1103/PhysRevA.82.031801)


\bibitem{variousstonepapers1} Chong, Y. D., Ge, L. \&  Stone, A. D. 2011
    PT-Symmetry Breaking and Laser-Absorber Modes in Optical Scattering
    Systems. \emph{Phys. Rev. Lett.} \textbf{106}, 093902.
    (doi:10.1103/PhysRevLett.106.093902)

\bibitem{variousstonepapers2}
Ge, L., Chong, Y. D. \& Stone, A. D. 2012
Conservation relations and anisotropic transmission resonances in one-dimensional PT-symmetric photonic heterostructures.
\emph{Phys. Rev. A} \textbf{85}, 023802.
(doi:10.1103/PhysRevA.85.023802)


\bibitem{yoo}
Yoo, G., Sim, H.-S.  \& Schomerus,  H. 2011
Quantum noise and mode nonorthogonality
in non-Hermitian PT-symmetric optical resonators.
\emph{Phys. Rev. A} \textbf{84}, 063833.
(doi:10.1103/PhysRevA.84.063833)

\bibitem{schomerusyh}
Schomerus, H and Yunger Halpern, N. 2013
Parity anomaly and Landau-level lasing in strained photonic honeycomb lattices.
\emph{Phys. Rev. Lett.} \textbf{110}, 013903.
(doi:10.1103/PhysRevLett.110.013903)


\bibitem{vahala}
Vahala, K. J. 2003
Optical microcavities.
\emph{Nature} \textbf{424}, 839--846.
(doi:10.1038/nature01939)

\bibitem{beenakkerreview}
Beenakker, C. W. J. 1997
Random-matrix theory of quantum transport.
\emph{Rev. Mod. Phys.} \textbf{69}, 731--808. (doi:10.1103/RevModPhys.69.731)

\bibitem{datta}
Datta, S. 1997 \emph{Electronic Transport in Mesoscopic Systems}.
Cambridge, UK: Cambridge University Press.

\bibitem{altlandzirnbauer}
Altland, A. \& Zirnbauer, M. R. 1997
Nonstandard symmetry classes in mesoscopic normal-superconducting hybrid structures.
\emph{Phys. Rev. B} \textbf{55}, 1142--1161. (doi:10.1103/PhysRevB.55.1142)


\bibitem{haake}
Haake,  F.  2010
\emph{Quantum signatures of chaos}, 3rd ed.
Berlin: Springer.

\bibitem{fyodorovsommers}
Fyodorov, Y. V. \& Sommers, H.-J. 1997
Statistics of resonance poles, phase shifts and time delays in quantum chaotic scattering: Random matrix approach for systems with broken time-reversal invariance.
\emph{J. Math. Phys.} \textbf{38}, 1918--1981.
(doi:10.1063/1.531919)

\bibitem{onsager} Onsager, L. 1931 Reciprocal relations in irreversible
    processes.  I. \emph{Phys. Rev.} \textbf{37}, 405.
    (doi:10.1103/PhysRev.37.405)

\bibitem{hsrmt}
Schomerus, H. 2011
Universal routes to spontaneous PT-symmetry breaking in non-Hermitian quantum systems.
\emph{Phys. Rev. A} \textbf{83}, 030101(R).
(doi:10.1103/PhysRevA.83.030101)

\bibitem{birchallweyl}
Birchall, C. \& Schomerus, H. 2012
Fractal Weyl laws for amplified states in PT-symmetric resonators.
arxiv:1208.2259

\bibitem{birchallrmt}
Birchall, C. \& Schomerus, H. 2012
Random-matrix theory of amplifying and absorbing resonators with PT or PTT$'$ symmetry.
preprint.

\bibitem{cannata}
Cannata, F.,  Dedonder, J.-P. \& Ventura, A. 2007
Scattering in PT-symmetric quantum mechanics.
\emph{Ann. Phys. (N.Y.)} \textbf{322}, 397--433.
(doi:10.1016/j.aop.2006.05.011)

\bibitem{berry}
Berry, M. V. 2008
Optical lattices with PT symmetry are not transparent.
\emph{J. Phys. A: Math. Theor.} \textbf{41}, 244007.
(doi:10.1088/1751-8113/41/24/244007)

\bibitem{jones}
Jones, H. F. 2007 Scattering from localized non-Hermitian potentials.
\emph{Phys. Rev. D} \textbf{76}, 125003. (doi:10.1103/PhysRevD.76.125003)

\bibitem{magnea}
Magnea, U. 2008
Random matrices beyond the Cartan classification.
\emph{J. Phys. A: Math. Theor.} {\bf 41}, 045203
(doi:10.1088/1751-8113/41/4/045203)

\bibitem{melsen} Melsen, J. A., Brouwer, P. W., Frahm, K. M. \&
    Beenakker, C. W. J. 1996
Induced superconductivity distinguishes chaotic from integrable
billiards. \emph{Europhys. Lett.} \textbf{35}, 7.
(doi:10.1209/epl/i1996-00522-9)

\bibitem{stonecpa1}
Chong, Y. D., Ge, L., Cao, H. \& Stone,  A. D. 2010
Coherent Perfect Absorbers: Time-Reversed Lasers.
\emph{Phys. Rev. Lett.} \textbf{105}, 053901.
(doi:10.1103/PhysRevLett.105.053901)

\bibitem{fyodsomeffs}
Fyodorov, Y. V. \&  Sommers, H.-J. 2000
Spectra of random contractions and scattering theory for discrete-time systems.
\emph{JETP Lett.} \textbf{72}, 422--426.
(doi:10.1134/1.1335121)

\bibitem{andreev}
Jacquod, Ph., Schomerus,  H. \&  Beenakker, C. W. J. 2003
Quantum Andreev Map: A Paradigm of Quantum Chaos in Superconductivity.
\emph{Phys. Rev. Lett.} \textbf{90}, 207004.
(doi:10.1103/PhysRevLett.90.207004)

\bibitem{schomerusjacquod} Schomerus, H. \& Jacquod, P. 2005
    Quantum-to-classical
    correspondence in open chaotic systems. \emph{J. Phys. A: Math. Gen.}
    \textbf{38}, 10663-10682. (doi:10.1088/0305-4470/38/49/013)

\bibitem{selfinverse2}
Haake,  F., Ku\'s, M., Sommers, H.-J.,  Schomerus, H. \& \.Zyczkowski, K. 1996
Secular determinants of random unitary matrices.
\emph{J. Phys. A: Math. Gen.} \textbf{29}, 3641--3658.
(doi:10.1088/0305-4470/29/13/029)

\bibitem{altlandthouless}
Altland, A., Gefen, Y. \& Montambaux, G. 1996
What is the Thouless Energy for Ballistic Systems?
\emph{Phys. Rev. Lett.} \textbf{76}, 1130--1133.
(doi:10.1103/PhysRevLett.76.1130)

\bibitem{mehta}
Mehta,  M. L. 2004 \emph{Random Matrices}, 3rd ed. New York, NY: Elsevier.

\bibitem{fractalweyl1}
Lu, W. T., Sridhar, S. \& Zworski, M. 2003
Fractal Weyl Laws for Chaotic Open Systems.
\emph{Phys. Rev. Lett.} \textbf{91}, 154101.
(doi:10.1103/PhysRevLett.91.154101)

\bibitem{fractalweyl3}
Schomerus, H. \& Tworzyd{\l}o,  J. 2004
Quantum-to-Classical Crossover of Quasibound States in Open Quantum Systems.
\emph{Phys. Rev. Lett.} \textbf{93}, 154102.
(doi:10.1103/PhysRevLett.93.154102)

\bibitem{west}
West, C. T., Kottos, T. \& Prosen,  T. 2010
PT-Symmetric Wave Chaos.
\emph{Phys. Rev. Lett.} \textbf{104}, 054102.
(doi:10.1103/PhysRevLett.104.054102)

\bibitem{beenakkerphase}  Paasschens, J. C. J.,  de Jong, M. J. M.,
    Brouwer, P. W.  \& Beenakker, C. W. J. 1997
Reflection of light from a disordered medium backed by a
phase-conjugating mirror.
    \emph{Phys. Rev. A} \textbf{56}, 4216--4228.
    (doi:10.1103/PhysRevA.56.4216)


\end{thebibliography}
\end{document}